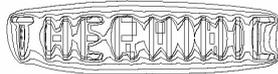

# THERMOS³, A TOOL FOR 3D ELECTROTHERMAL SIMULATION OF SMART POWER MOSFETS

*Giovanni Buonaiuto, Andrea Irace, Giovanni Breglio, Paolo Spirito*

University of Naples Federico II,
Department of Electronic and Telecommunication Engineering, Italy.

**ABSTRACT**

In this work we present a novel 3D simulation tool capable of taking into account also particular driving strategies of the electron device as it may be the case of Smart Power MOSFETs where a control logic interacts with the power section and controls its dissipated power and temperature. As an example a thermal shutdown circuit, capable of reading the temperature on chip and switching the device off if the latter reaches dangerous values is usually embedded within Smart Power devices used in automotive applications to drive direction light or small motors/actuators.

## 1. INTRODUCTION

The simulation of power electron devices is a complex numerical problem and it has been faced in many different ways in the recent past as the issue of power dissipation together with the knowledge of electro-thermal interactions in MOSFETs has become relevant. Being coupled, but on a different timescale with respect to the thermal problem, the electrical problem is usually reduced to few DC equation that are sufficient to model the temperature dependent static behavior of the single cell. The time-dependent heat equation is solved, numerically or analytically, in a way such as to take into account layout geometries, boundary conditions and heatsink influence.

The THERMOS³ simulator we propose, entirely written in MATLAB, is based on a forward iterative finite difference time domain scheme [1] where the electrical equivalent of the thermal problem and the electrical quantities (drain current, source voltages) are solved self-consistently at each discrete time step. Non linearity in the thermal conductivity together with voltage depolarization due to finite resistance of the source metallization is taken into account. The $I_D=f(V_{GS},V_{DS},T)$ behavior of the unit cell is modeled considering temperature dependence of the threshold voltage, mobility and MOSFET internal resistances. Convection on the surface and sides and isothermal boundary condition at the heatsink have been used.

As an exhaustive example of the capability of this new tool we present the simulation of a smart power device produced by STMicroeletronics (Fig.1). This device is protected during permanent short circuit by a current feedback loop which keeps the dissipated power inside the SOA together with a thermal shutdown circuit which shuts the device of if temperature on chip exceeds a reference value. Dynamic temperature behaviour of this device has been fully characterized by fast transient infrared imaging [2] therefore a quantitative comparison with the simulation is available for validation. One main issue is the optimization of timestepping since the overall behaviour of the device has to be observed in the 100ms (or longer) timescale while sensitive information such as fast temperature transients can be as short as 100μs. In our approach an adaptive timstepping strategy together with switching event prediction and detection has been implemented.

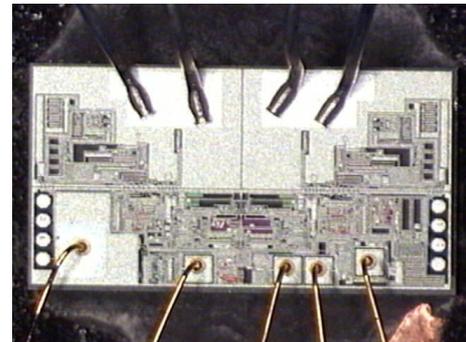

Fig. 1 Optical picture of the two channels monolithic smart Power MOSFET implemented with M0 technology (VND830SP).

## 2. THE SIMULATOR

A multicellular power VDMOS can be treated as a number of equivalent single cells all of them sharing the same voltage on the drain and gate terminal and with the sources connected through a metal layer with a finite resistivity. Therefore both $V_{DS}$ and $V_{GS}$ are dependent not only on temperature (the resistivity of the metal layer is itself a function of temperature) but on the layout of the device and position of the source bond wires. The single cell is modelled by the following set of equations which are a version of the SPICE LEVEL 1 model modified to take explicitly into account temperature dependence of the threshold voltage and mobility. All the internal capacitances are neglected because the transient problem is observed on a different timescale, and also the other devices (drain-body diode etc.) which are usually considered in power MOSFET models since we are interested in the behaviour of the device when it is driven in its triode or pinch-off region. Therefore for the generic cell we have:

$$V_T = V_{T0} - \varphi(T - T_0)$$
$$K(T) = \frac{1}{2}\mu(T)C_{ox}\frac{W}{L} \quad (1)$$
$$\mu(T) = \mu_0 \left(\frac{T}{T_0}\right)^{-m}$$

$$I_D = \begin{cases} 0 \text{ if } V_{GS} \leq V_T \\ K\left[2(V_{GS}-V_T)V_{DS} - V_{DS}^2\right] \text{ if } V_{GS} \geq V_T \text{ and } V_{DS} - V_{GS} \leq V_T \\ K(V_{GS}-V_T)^2 \text{ if } V_{GS} \geq V_T \text{ and } V_{DS} - V_{GS} \geq V_T \end{cases} \quad (2)$$

Depolarization effects due to finite resistance of the source metal layer can be taken into account by solving the linear problem:

$$\overline{\overline{Y_E}}(T) \cdot \overline{V} = \overline{I}(T) \quad (3)$$

where $Y_E$ is the admittance matrix (where temperature dependence of the metal resistivity is taken into account) which defines the electrical network modelling the metal layer, V is the vector of the voltages at each node and I are the current flowing into each node (that is the drain current of the active MOSFET cells).

Given the temperature field $T(x,y)$ on the surface of our domain, both $V_{DS}$ and $I_D$ are known at each cell and therefore it is know the dissipated power at the particular location, that is the heat flux field $Q(x,y)$ according to the equations:

$$Q(x,y) = \frac{V_{DS}(x,y)I_D(x,y)}{A} \quad (4)$$

Since we discretize the domain in x and y directions the equation (4) reduces to its discrete form which gives us the discrete heat sources needed to solve the thermal problem

$$Q_{i,j} = \frac{V_{i,j}I_{i,j}}{A_{i,j}} \quad (5)$$

The power device is treated as a surface heat source, being the heat flux equal to the power dissipated by joule heating within the device, while the solder and heat sink are modelled through simple RC networks. Convection/Radiation from the top and side surfaces are also taken into account.

By writing heat balance at each node of the domain we can completely describe our problem by its matrix form

$$\overline{\overline{Y_T}}(T) \cdot \overline{T} = \overline{Q}(T) \quad (6)$$

where the admittance matrix $Y_T$ is a sparsely populated matrix, T is the vector which defines temperatures at each node of the domain and Q is the vector which takes into account the heat flux injected into the domain as a consequence of the dissipated power by the power device. The solution strategy we propose, entirely implemented in MATLAB, is based on a forward iterative finite difference time domain scheme [3] where the electrical equivalent of the thermal problem as described by eq.4 and the electrical quantities (drain current, source voltages) of eq.3 are solved iteratively at each discrete time step until a number desired convergence criteria is reached. Before entering the details of the simulation results few words regarding driving strategies of Smart Power Devices and consequent meshing and time stepping solutions have to be spent.

In our case the single unit cell is too small to be considered as a good elementary cell to discretize our domain. Therefore we decide to group a number of single devices and treat them as a macro cell with the current scaled by its area. We usually start with a 50μm x 50μm cell although, during the transient simulation, an adaptive remeshing strategy is used to refine the mesh where the temperature gradients are higher. In the case of the VND800PEP, which is about 2mm x 2mm the starting grid results in 40x40 elementary cells. During the transient solution the number of nodes is kept constant whereas their location is moved in the region where temperature gradients are higher.

Regarding time stepping strategy, the issue is slightly different. We have already described the driving strategy of this kind of devices and, in general, as we will deal with smart power devices, where dynamics can happen with very different dynamics and timescales, a fixed time step strategy can increase the computational time to non practicable solutions. We therefore choose an adaptive time stepping strategy where the time step shortens as temperature and voltage derivatives increase. This on its turn poses a problem when the control logic switches the power on and off being at this time instant the $V_{GS}$ derivative is theoretically infinite demanding for a time step which approaches to zero. We choose in this case an event-driven predictive algorithm in order to foresee the instant when the condition for commutation will be verified and adjust the time step accordingly.

This is a quite new approach in the solution of electrical network with a high number of switches and it can be also implemented in this case where all the unit cells can be treated as a single switch.

### 3. THE FAST TRANSIENT INFRARED IMAGING

The detection of the temperature distribution across the area of the device in transient conditions has been made possible by the use of a custom developed radiometric 2D measurement system that allows acquiring the transient temperature maps of the device surface [2]. We use a fast (actual time resolution is less than 2 µs) single cooled Cd-Mg sensor mounted in a microscope optical chain that has an equivalent spot size of about 10 µm with a working distance of 2.5 mm. In order to perform a 2D scanning of the surface to be mapped, we use an x-y step motor stage controlled via a motor control card installed in a PC which controls device biasing and data acquisition. In such way, the device under test is shifted with respect to the microscope spot, and it is possible to cover the entire surface under observation.

The step by step acquisition of the radiometric waveforms coming from the sensor is performed by means of a 5MS/s 12bit A/D converter card that connected via the PCI bus into the PC converts and elaborates the output detected signal. The acquisition software also performs the pre-filtering of the detected signal by using an averaging procedure to increase the signal to noise ratio.

It is important to remember that radiometric systems are able to perform absolute thermal measurement only if the target emissivity coefficient is known. Unfortunately, the surface of power devices is often composed of different materials (i.e. aluminum, passivation, silicon oxide, etc.) that are characterized by different emissivity coefficients. Hence, in our thermal mapping procedure, a preliminary characterization step is also performed in order to determine the emissivity coefficients of the points that define the acquisition grid.

The temperature map is reconstructed by acquiring the emissivity signals synchronized with the electrically switching and after that the numerical processing of the radiometric transient signals starts.

First of all, the stored dynamic radiometric signals are properly corrected by using the emissivity coefficient of the corresponding grid point and converted into the true dynamic temperature signal. Then, the temperature array data manipulated to obtain the time frames of the temperature distribution on the DUT.

By means of this equipment and following the previously described procedure we have obtained the thermal transient analyses used to optimize the performances of the devices described in this work.

### 4. THE VDN SMARTH POWER MOS AND ITS PROTECTION STRATEGY

Figure 1 shows a smart power MOSFET made with M0TM technology integrating a DMOS and its control part in a monolithic chip solution. To comply with automotive requirements this class of devices includes several protections that are described in fig. 2.

#### 4.1. Short circuit protection

The more restricting rule driving the protection strategy of the Power actuator is to avoid false short circuit detections in the harsh and very noisy automotive environment. Beside a high level of electromagnetic susceptibility the devices should not turn off either for an inrush condition either for intermittent short circuit. The short circuit protection is implemented using only negative feedback method to ensure stability and to maintain the device in a predictable status in every condition.

A Current limitation combined with thermal shut down intervention protects the power device during short circuit operation. The integrated current feedback fixes the working point in the active area of the Power stage.

The resulting high power dissipated equal to battery voltage multiplied by Drain current leads to a fast increase of the thermal sensor temperature and to the intervention of thermal shut down block.

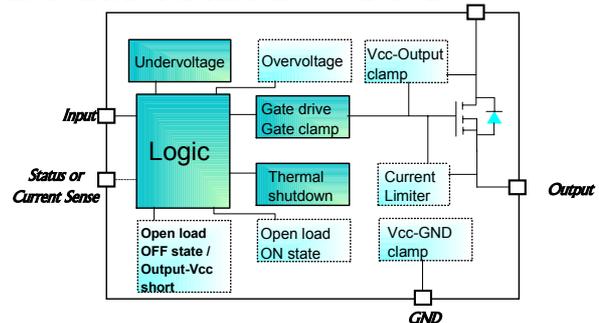

Fig. 2. Functional block diagram of the smart Power MOSFET

#### 4.2 Current density versus technology evolution

The balance of power dissipation due to joule effect ($P_d = R_{on} * I^2$) and the ability to extract the heat ($P_d = R_{TH} * (T_J - T_{AMB})$) establish the current rating of a Power device. In actual automotive junction boxes Smart Power are surface mounted and the heat is dissipated via the cupper traces of a PCB in FR4. With this boundary condition the die size reduction does not affect significantly the thermal resistance junction ambient so it is the on resistance of the device that manly sets the current rating.

Scaled technology permitted a significant die size reduction and the $R_{on}*mm^2$ has decreased by a factor 10 in the last 10 years. As a result the size of the silicon can be reduced significantly for the same targeted load.

On the other hand the short circuit current of the device cannot be reduced because this value needs to be higher than the inrush current of the targeted loads. The result is that also the power density generated in the device during short circuit increased also by a factor 10.

The protection strategy is very simple (Fig.3): as temperature at sensor location reaches a high threshold value (Tshutdown) drain current is switched off and the device is left to its free thermal evolution; when temperature then reaches a low threshold value (Treset) drain current is switched on again. In this way power is controlled, the device is kept well inside the SOA and failures are prevented (although reliability problems due to metal fatigue can occur [2]).

At this point the temperature is toggling between max temperature fixed and fixed temperature reset. The device returns automatically to nominal operation when the short circuit disappears so the device continues to operate properly.

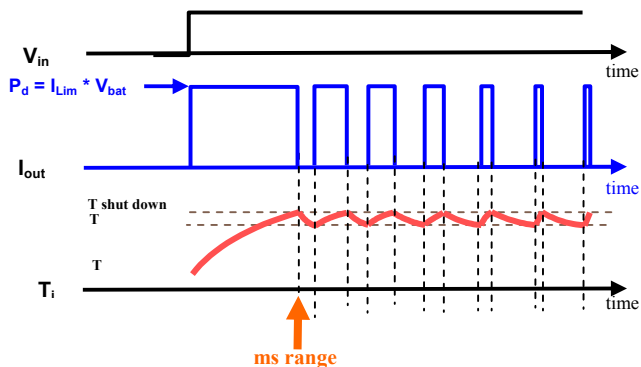

Fig. 3 The sketched representation of the protection strategy of the device under short circuit conditions: Time plots in milliseconds scale of the whole dissipated power and temperature of the embedded thermal sensor.

### 6. RESULTS, COMPARISONS AND COMMENTS.

The device has been discretized in $Nx \times Ny \times Nz$ unit cells each with its own thermal parameters and operated in short circuit to the usual battery voltage Vbat=12V. Current feedback fixes the total current at Ilim=10A. In Fig.4 we report a comparison between experimentally detected temperature distribution and the result of the simulation. As we can note a good quantitative agreement is obtained both in shape and in peak temperature value.

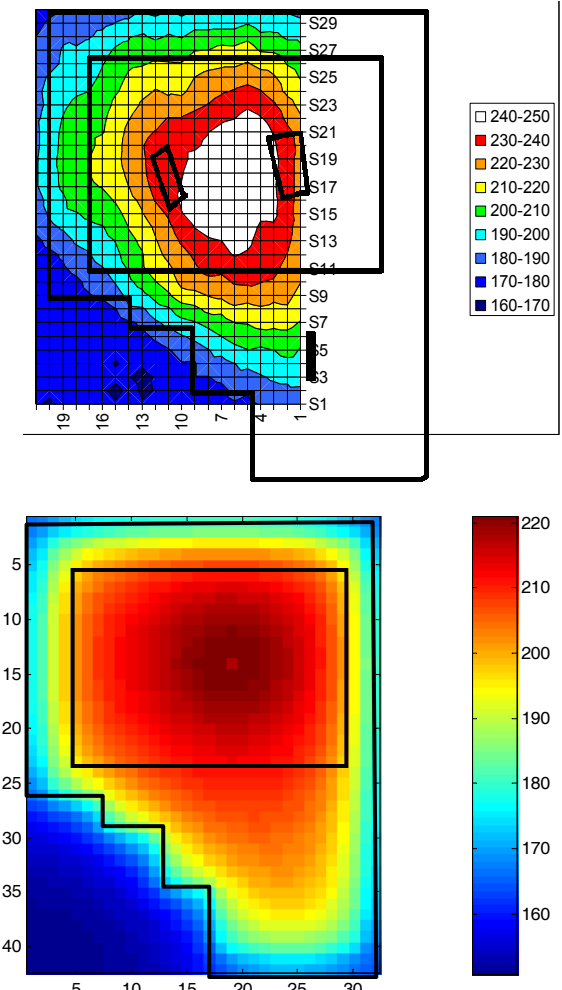

Fig.4: Temperature distribution on chip: experimental (up) and simulated (down)

To further investigate the features of the simulator we would like to underline that a well designed thermal shutdown strategy relies on the optimal position of the thermal sensor on the device surface. If this should not be the case, temperature difference between the sensor and the maximum temperature reached on chip can be observed and reliability of the device can be eventually impaired as temperature fatigue of the metals and soldering joint might increase the Ron of the device. In Fig.5 we report a simulation where the temperature sensor has been deliberately placed far from the region where temperature reaches its maximum, we notice how, even if the sensor is correctly switching the device, the maximum temperature on chip reaches values higher than Tshutdown.
From Fig.5 and Fig. 6 the PWM modulation of the dissipated power and the timestepping strategy can also be observed. In particular (see Fig. 6) it is noticeable that as temperature at the sensor location reaches the shutdown threshold timestepping is reduced thanks to the predictive event detection algorithm and the same happens when temperature reaches the reset value with a negative slope.

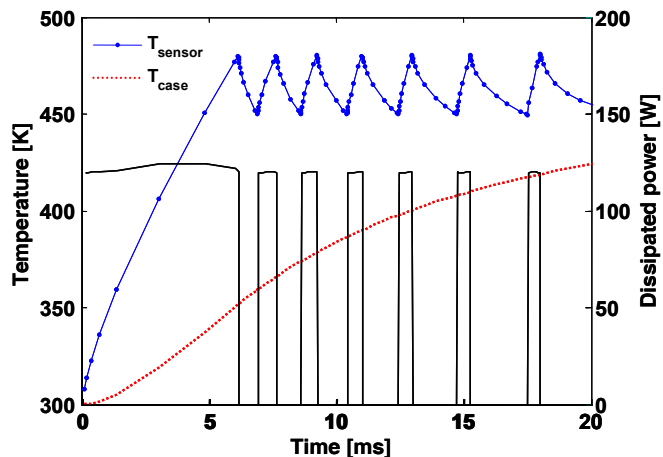

Fig.5 Simulation of the permanent short circuit protection on a VND800PEP device

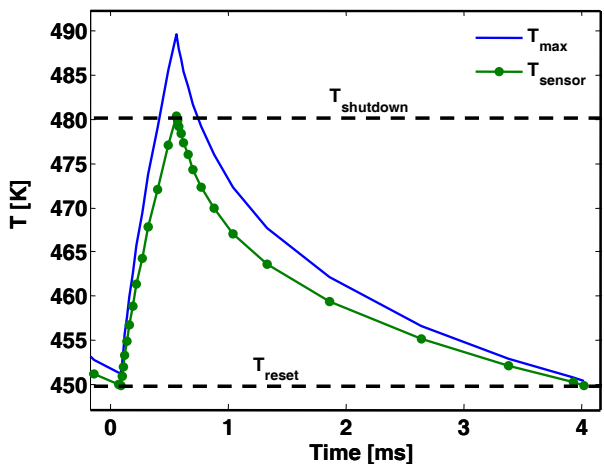

Fig. 6 A detail of a single heating-cooling transient with comparison between the maximum temperature on chip and temperature detected by the thermal sensor

The value of such a tool in the design stage of a smart power device is straightforward as it can be used to optimize the whole design, influence of layout (i.e. position of temperature sensors, bond pads location etc.) and to investigate the performances of different protection strategies.

## 7. FURTHER IMPLEMENTATION

In our study on these kind of devices and in all the experienced acquired in the electro-thermal simulation of power devices, we have understood that not only the localization of the bond wires is important in simulations but also a proper electro-thermal description of these components must to be taken into account.
Into the next new version of the THERMOS³ simulation tool will be inserted also the possibility to well simulate and describe how the power bond wires, described in terms of position contact area with the device, thick and length, of power MOS can affect the electro-thermal behavior of the whole device.

## 8. REFERENCES

[1] G. Breglio and P. Spirito "Experimental detection of time dependent temperature maps in power bipolar transistors", Microelectronics Journal, Volume 31, Issue 9-10, October 2000, pp. 735-739

[2] A. Irace, G. Breglio, P. Spirito, R. Letor, S. Russo, "Reliability enhancement with the aid of transient infrared thermal analysis of smart Power MOSFETs during short circuit operation", Microelectronics Reliability, Volume: 45, Issue: 9-11, September - May, 2005, pp. 1706-1710

[3] A.F. Mills, *Heat and mass transfer*, Richard D. Irwin ed.,1995.